\documentclass{aastex}
\usepackage{spr-astr-addons}
\usepackage{url}
\usepackage{graphicx}
\usepackage{epsfig}
\usepackage{epstopdf}

\RequirePackage{color}

\begin{document}

\title{Gamow-Teller strength distributions and neutrino energy loss rates due to  chromium isotopes in stellar matter\\}

\shorttitle{GT and neutrino cooling rate calculations due to Cr
isotopes} \shortauthors{Nabi et al.}

\author{Jameel-Un Nabi\altaffilmark{1}} \and \author{Ramoona Shehzadi\altaffilmark{2}} \and
\author{Muhammad Fayaz\altaffilmark{1}}

\email{jameel@giki.edu.pk}

\altaffiltext{1}{Faculty of Engineering Sciences,\\GIK Institute of
Engineering Sciences and Technology, Topi 23640, Khyber Pakhtunkhwa,
Pakistan.} \altaffiltext{1}{Corresponding author email :
jameel@giki.edu.pk} \altaffiltext{2}{Department of Physics,
University of the Punjab, Lahore, Pakistan.}

\begin{abstract}
Gamow-Teller transitions in isotopes of chromium play a
consequential role in the presupernova evolution of massive stars.
$\beta$-decay and electron capture rates on chromium isotopes
significantly affect the time rate of change of lepton fraction
($\dot{Y_{e}}$). Fine-tuning of this parameter is one of the key for
simulating a successful supernova explosion. The (anti)neutrinos
produced as a result of electron capture and $\beta$-decay are
transparent to stellar matter during presupernova phases. They carry
away energy and this result in cooling the stellar core. In this
paper we present the calculations of Gamow-Teller strength
distributions and (anti)neutrino energy loss rates due to  weak
interactions on chromium isotopes of astrophysical importance. We
compare our results with measured data and previous calculations
wherever available.
\end{abstract}


\section{Introduction}
Elements heavier than helium are synthesized in the stars via fusion
chain reactions during the course of stellar evolution. Once the
nucleosynthesis process cooks iron ($^{56}$Fe), the binding energy
per nucleon curve prohibits any further production of energy by
nuclear fusion. When the iron core, formed in the center of a
massive star, grows by silicon shell burning to a mass around
1.5\;M$_\odot$, the electron degeneracy pressure required to counter
the enormous self-gravity force of the star is reduced and the core
becomes unstable. This starts what is normally termed as
core-collapse supernova, in the course of which the star explodes
and the parts of heavy-element core and its outer shells are ejected
into the intersteller medium. In addition to disseminating nuclei
formed during stellar evolution, supernova play a key role in
synthesizing additional heavy elements. These elements (heavier than
iron) are thought to be formed in the r-process nucleosynthesis and
type II supernova explosions are considered to be the most probable
site for such neutron-capture processes \citep{Cow04}. Thus, a
detailed understanding of the explosion mechanism is not only
necessary for the fate of a star's life but also for understanding
the very important nucleosynthesis problem which is intricately
linked with the microphysics of core collapse.

Since 1934, when Baade and Zwicky~\citep{Baade34} proposed that a
supernova represents the transition of a normal star to a neutron
star, scientists have been debating about the details of the
mechanism responsible for these spectacular explosions.
Unfortunately, the supernova explosion mechanism is still
mysterious. To date, core- collapse simulators find it challenging
to successfully transform the collapse into an explosion.

Core collapse creates high temperatures ($>$ 1\;MeV) and densities
($10^{7}$\;g\;cm$^{-3} < \rho < $ 10$^{15}$\;g\;cm$^{-3}$) and
produces either a neutron star or black hole. Under such extreme
thermodynamic conditions, neutrinos are produced in abundance. The
discovery of neutrino burst from SN1987A by the Kamiokande II group
\citep{Hir87} and IMB group \citep{Bio87} energized the research on
neutrino astrophysics. Neutrinos are considered to play a critical
role in our understanding of the dynamics of supernova. They not
only provide an essential probe into the core-collapse mechanism,
but also probably play an active role in the explosion mechanism.
They seem to be mediators of the transfer of energy from the inner
core to the outer mantle of the iron core. This energy transfer is
thought to transform the collapse into an explosion.

According to our present understanding of the explosion mechanism,
the onset of the collapse and infall dynamics are very sensitive to
the core entropy and to the lepton to baryon ratio,
Y$_e$~\citep{Bethe79}. These two quantities are mainly determined by
weak interaction processes, electron capture and $\beta$-decay.
Electron capture on (Fe-peak) nuclei is energetically favorable when
electron Fermi energy reaches the nuclear energy scale (energies in
the MeV range). This produces neutrinos and reduces the number of
electrons available for the pressure support. Many of the nuclei
present can also $\beta$-decay, a process which acts in the opposite
direction. These weak-interaction mediated reactions affect the
overall lepton to baryon ratio of the core and directly influence
the collapse dynamics.

At the start of collapse, while densities in the core are $\sim
10^{9}$\;g\;cm$^{-3}$, the neutrinos can freely escape the
collapsing star and therefore assist in cooling the core to a lower
entropy state. As the core density reaches $\rho \sim
10^{12}$\;g\;cm$^{-3}$, an important change occurs in the physics of
the core collapse. At such high densities, (where neutrino diffusion
time becomes larger than the collapse time)~\citep{Bethe90}
neutrinos feel the pressure of matter and can become trapped in the
so-called neutrinosphere, mainly due to elastic scattering with
nuclei. After neutrino trapping, the collapse proceeds essentially
homologously~\citep{Goldreich80}, until nuclear densities, $\rho
\sim 10^{14}$\;g\;cm$^{-3}$, are reached. The homologous core
decelerates and bounces in response to the increased nuclear matter
pressure. This drives a shock wave into the outer core. If the shock
were to propagate outward without stalling, we would have what is
called prompt-shock mechanism. However, it appears as if the energy
available to the shock is not sufficient, as it loses energy in
dissociating heavy nuclei that pass through it as it propagates
outward. The shock loses additional energy due to neutrino emission
(mainly non-thermal) leading to a standing shock. This stalled shock
is thought to be revived by what has become known as the
delayed-shock mechanism, originally proposed by Wilson and
Bethe~\citep{Bethe85}. The success of this explosion mechanism
mainly depends on the cross sections for neutrino capture on
nucleons, the neutrino production rates, the details of the neutrino
transport, neutrino cooling rates, convection behind the stalled
shock, and other factors, many of which are not known with
certainty. Depending on these ingredients, simulations of the
delayed shock mechanism yielded successful explosions in some
cases~\citep{Herant94, Fryer00}, while they failed in some of the
other cases~\citep{Buras03}-\citep{Thompson03}.

After core bounce, $\sim 10^{53}$\;ergs of
energy~\citep{Balantekin03} is released from the cooling
proto-neutron star in the form of neutrinos and antineutrinos of all
3 flavours (electron, muon and tau). Only $1\%$ of this energy needs
to be absorbed behind the shock to generate the $\sim 10^{51}$\;ergs
of energy associated with the explosion. Simulating a $1\%$ effect
is indeed a challenging task faced by astrophysicists all around the
globe. Neutrinos while propagating through the proto-neutron star
interact with the protons, neutrons and electrons in this central
object via absorption and scattering. These neutrinos play a vital
role in our understanding of the microphysics of the supernova. They
provide information concerning the neutronization due to electron
capture, the formation, stalling, revival and propagation of the
shock wave and the cooling phase. Electron capture rates and
associated neutrino cooling rates (as a function of stellar
temperature, density and Fermi energy) are important input
parameters for modeling the dynamics and supernova collapse phase of
massive stars~\citep{Strother09}. A reliable and microscopic
calculation of neutrino loss rates and capture rates has thus become
a topic of interest for core-collapse simulators all around the
world.

Fuller, Fowler, and Newman (FFN)~\citep{Fuller} were the first who
performed a comprehensive calculation of stellar weak rates
including the capture rates, neutrino energy loss rates and decay
rates for a wide range of density and temperature. The calculation
was done for 226 nuclei in the mass range $21 \le A \le 60$. FFN
work was later extended by Aufderheide et al. \citep{Aufderheide94}
for heavier nuclei with $A > 60$. Since then theoretical efforts
were focused on the microscopic calculations of capture rates of
iron-regime nuclide. The proton-neutron quasi particle random phase
approximation theory (pn-QRPA) \citep{Nabi99} and large-scale shell
model \citep{Lan00}  were used largely for the microscopic
calculation of weak interaction rates in stellar environment.

Weak interaction rates for 709 nuclei with A = 18 to 100 in stellar
matter using the pn-QRPA theory were calculated by Nabi and Klapdor
~\citep{Nabi99}. These weak interaction rate calculations included
decay rates, capture rates, neutrino energy loss rates, gamma
heating rates, probabilities of $\beta$-delayed particle emissions
and energy rate of these particle emissions~\citep{Nabi99a, Nabi04}.
These calculations were later refined using more efficient
algorithms, incorporating latest data from mass compilations and
experimental values, and by fine-tuning of model
parameters~\citep{Nabi05}-\citep{Nabiii07}. The present work is
devoted to a microscopic calculation of Gamow-Teller strength
distribution and detailed analysis of the neutrino and antineutrino
energy loss rates due to weak-interaction reactions on isotopes of
chromium in stellar environment. The neutrino and antineutrino
energy loss rates can occur through four different weak-interaction
mediated channels: electron and positron captures, and, electron and
positron emissions. The neutrinos are produced due to electron
captures and positron decays whereas the antineutrinos are produced
due to positron captures and electron decays.

Charge-changing transitions normally referred to as Gamow- Teller
(GT) transitions play an important role in many astrophysical events
in universe. During early stages of collapse many important nuclear
processes, such as $\beta$ decays, electron captures, neutrino
absorption and inelastic scattering on nuclei, appear. These
reactions are mainly governed by GT (and Fermi) transitions. GT
transitions for $fp$-shell nuclei are considered very important for
supernova physics \citep{Fuller}. The GT transitions, in $fp$-shell
nuclei, play decisive roles in presupernova phases of massive stars
and also during the core collapse stages of supernovae (specially in
neutrino induced processes). At stellar densities $\sim$ 10$^{11}$
gcm$^{-3}$, for $fp$-shell nuclei, the electron chemical potential
approaches the same order of magnitude as the nuclear Q-value. Under
such conditions, the $\beta$-decay rates are sensitive to the
detailed GT distributions. For still higher stellar densities, the
electron chemical potential is much larger than nuclear Q-values.
Electron capture rates become more sensitive to the total GT
strength for such high densities. To achieve a better understanding
of these notoriously complex astrophysical phenomena, a microscopic
calculation of  GT strength distributions is in order.

The GT excitations deal with the spin-isospin degree of freedom and
are executed by the $\sigma\tau_{\pm,0}$ operator, where $\sigma$ is
the spin operator and $\tau_{\pm,0}$ is the isospin operator in
spherical coordinates. The plus sign refers to the GT$_+$
transitions where a proton is changed into a neutron (commonly
referred to as electron capture or positron decay). On the other
hand, the minus sign refers to GT$_-$ transitions in which a neutron
is transformed into a proton ($\beta$-decay or positron capture).
The third component GT$_{0}$ is of relevance to inelastic
neutrino-nucleus scattering for low neutrino energies and would not
be considered further in this manuscript. Total GT$_-$ and GT$_+$
strengths (referred to as $B(GT)_-$ and $B(GT)_+$, respectively, in
this manuscript) are related by Ikeda Sum Rule as
$B(GT)_{-}-B(GT)_{+}=3(N-Z)$, where $N$ and $Z$ are numbers of
neutrons and protons, respectively \citep{Ike63}. Given nucleons are
treated as point particles and two-body currents are not considered,
the model independent Ikeda Sum Rule should be satisfied by all
calculations.

Isotopes of chromium  are advocated to play an important role in the
presupernova evolution of massive stars. The measured data of GT
strength in chromium isotopes have been sparse to the best of our
knowledge. The decay of $^{46}$Cr was studied by \citep{Zio72}, who
used the $^{32}$S($^{16}$O,2n) reaction to produce $^{46}$Cr. Later
Onishi and collaborators \citep{On05} observed the $\beta$-decay of
$^{46}$Cr to the 1$^{+}_{1}$ state  at 993 keV excitation energy in
$^{46}$V. The $T = 1$ nuclei decay to the $T = 0$ and 1$^{+}$ states
of daughter nuclei were called favored-allowed GT transitions and
possessed a signature small $ft$ value. The experiment was performed
at RIKEN accelerator research facility. Two sets of independent
measurement of $B(GT)_-$ strength for $^{50}$Cr were also performed.
Fujita et al. did a $^{50}$Cr($^{3}$He, $t$)$^{50}$Mn measurement up
to 5 MeV in daughter \citep{Fuj11}. On the other hand Adachi and
collaborators \citep{Ada07} were able to perform a high resolution
$^{50}$Cr($^{3}$He, $t$)$^{50}$Mn measurement at an incident energy
of 140 MeV/nucleon and at 0$^{0}$ for the precise study of GT
transitions. The experiment was performed at RCNP, Japan. Owing to
high resolution the authors were able to measure $B(GT)_-$ strength
up to 12 MeV in $^{50}$Mn. At higher excitations above the proton
separation energy, a continuous spectrum caused by the quasifree
scattering appeared in the experiment.  Nonetheless there was a need
to perform more experiments to measure GT transitions in $fp$-shell
nuclei. Next-generation radioactive ion-beam facilities (e.g. FAIR
(Germany), FRIB (USA) and FRIB (Japan)) are expected to provide us
measured GT strength distribution of many more nuclei. It is also
expected to observe GT states in exotic nuclei near the neutron and
proton drip lines. However simulation of astrophysical events (e.g.
core-collapse supernovae) requires GT strength distributions ideally
for hundreds of nuclei. As such experiments alone cannot suffice and
one has to rely on reliable theoretical estimates for GT strength
distributions.

Aufderheide and collaborators \citep{Aufderheide94} searched for key
weak interaction nuclei in presupernova evolution of massive stars.
Phases of evolution, after core silicon burning, were considered and
a search was performed for the most important electron capture and
$\beta$-decay nuclei for 0.40 $\le Y_{e} \le$ 0.50 ($Y_{e}$ is
lepton-to-baryon fraction of the stellar matter). The rate of change
of $Y_{e}$ during presupernova evolution is one of the keys to
generate a successful explosion. As per their calculation, electron
captures on $^{51-58}$Cr and $\beta$-decay of
$^{53,54,55,56,57,59,60}$Cr were found to be of significant
astrophysical importance controlling $Y_{e}$ in stellar matter.
Later Heger and collaborators \citep{Heg01} performed simulation
studies of presupernova evolution employing shell model calculations
of weak-interaction rates in the mass range A = 45 to 65. Electron
capture  rates on $^{50,51,53}$Cr were found to be crucial for
decreasing the $Y_{e}$ of the stellar matter. Similarly, it was
shown in the same study that $\beta$-decay rates of
$^{53,54,55,56}$Cr played a significant role in increasing the
$Y_{e}$ content of the stellar matter.  These and similar studies of
presupernova evolution provided us the motivation to perform a
detailed study of GT transitions  and the (anti)neutrino energy loss
rates due to isotopes of chromium. In this work we calculate and
study GT distributions of eleven (11) isotopes of chromium,
$^{50-60}$Cr, both in the electron capture and $\beta$-decay
direction. The (anti)neutrino energy loss rates of these chromium
isotopes are also calculated and compared with previous
calculations.

The theoretical formalism used to calculate the GT strength
distributions and associated (anti)neutrino energy loss rates in the
pn-QRPA model is described briefly in the next section. We compare
our results with other model calculations and measurements in
Section 3. A decent comparison would put more weight in the
predictive power of the pn-QRPA model used in this work. It is
pertinent to mention again that our calculation includes many
neutron-rich and neutron-deficient isotopes of chromium for which no
experimental data is available for now. Core-collapse simulators
rely heavily on reliable theoretical estimates of the corresponding
weak rates in their codes. Section 4 finally summarizes our work and
states the main findings of this study.

\section{Model Selection and Formalism}
GT strength distributions of $^{50-60}$Cr isotopes were calculated
by using the proton-neutron quasiparticle random phase approximation
(pn-QRPA) model (we refer specially to the calculations by
\citep{Sta90, Hir93}). The QRPA treats the particle-particle ($pp$)
and hole-hole amplitudes in a similar way as in particle-hole ($ph$)
amplitudes. The QRPA takes into account pairing correlations albeit
in a non-perturbative way. Earlier a similar study  for calculation
of GT transitions for key chromium isotopes, using different pn-QRPA
models, was performed \citep{Cak15}. The idea was to find out the
best pn-QRPA model to perform the stellar weak interaction rates
with the respective model parameters. It was concluded in
\citep{Cak15} that the current pn-QRPA model was indeed the best
model that reproduced not only the available experimental data but
had the best predictive power for estimate of weak rates of nuclei
far away from line of stability. In this paper we use the same model
(with same model parameters) to calculate (anti)neutrino cooling
rates due to isotopes of chromium.

In this section, we give necessary formalism used in the pn-QRPA
models. Detailed formalism may be seen in \citep{Nabi04} and is not
reproduced here for space consideration.
\subsection{The GT Strength Distribution}
The Hamiltonian of the pn-QRPA model is given by
\begin{equation}
H^{QRPA} = H^{sp} + V^{pair} + V ^{ph}_{GT} + V^{pp}_{GT}.
\label{Eqt. 9}
\end{equation}
Single particle energies and wave functions were calculated in the
Nilsson model which takes into account nuclear deformation. Pairing
in nuclei was treated in the BCS approximation. In the pn-QRPA
model, proton-neutron residual interaction occurs through both $pp$
and $ph$ channels. Both the interaction terms were given a separable
form. $V_{GT}^{ph}$ is the particle-hole (ph) GT force, and
$V_{GT}^{pp}$  is the particle-particle (pp) GT force. The
proton-neutron residual interactions occurred as particle-hole and
particle-particle interaction. The interactions were given separable
form and were characterized by two interaction constants $\chi$  and
$\kappa$, respectively.
Other parameters required for the calculation of weak rates
are the Nilsson potential parameters, the pairing gaps, the
deformations, and the Q-values of the reactions. Nilsson-potential
parameters were taken from Ref. \citep{Nil55} and the Nilsson
oscillator constant was chosen as $\hbar \omega=41A^{-1/3}(MeV)$
(the same for protons and neutrons). The calculated half-lives
depend only weakly on the values of the pairing gaps \citep{Hir91}.
Thus, the traditional choice of $\Delta _{p} =\Delta _{n}
=12/\sqrt{A} (MeV)$ was applied in the present work. Experimentally
adopted values of the deformation parameters, for even-even isotopes
of chromium ($^{50,52,54}$Cr), extracted by relating the measured
energy of the first $2^{+}$ excited state with the quadrupole
deformation, were taken from \citep{Ram87}. For other cases the
deformation of the nucleus was calculated as
\begin{equation}
\delta = \frac{125(Q_{2})}{1.44 (Z) (A)^{2/3}},
\end{equation}
where $Z$ and $A$ are the atomic and mass numbers, respectively, and
$Q_{2}$ is the electric quadrupole moment taken from M\"{o}ller and
collaborators \citep{Moe81}. Q-values were taken from the recent
mass compilation of Audi and collaborators \citep{Aud12}. Our
ultimate goal is to calculate reliable and microscopic weak rates
for astrophysical environments, many of which cannot be measured
experimentally. The theoretical calculation poses a big challenge.
For example, it was concluded that $\beta$-decay and capture rates
are exponentially sensitive to the location of GT$_{+}$ resonance
while the total GT strength affect the stellar rates in a more or
less linear fashion \citep{Auf96}. Weak rates, with an excited
parent state, are required in sufficiently hot astrophysical
environments.

The results of pn-QRPA calculations were multiplied by a quenching
factor of $f_{q}^{2}$ = (0.6)$^{2}$ \citep{Vet89, Gaa83} in order to
compare them with experimental data and previous calculations, and
to later use them in calculation of (anti)neutrino energy loss
rates. Interestingly \cite{Vet89} and \cite{Roe93} predicted the
same quenching factor of 0.6 for the RPA calculation in the case of
$^{54}$Fe when comparing their measured strengths to RPA
calculations.

The Ikeda Sum Rule, in re-normalized form, in our model translates
to
\begin{equation}
ISR_{re-norm} = B(GT)_{-}-B(GT)_{+}\cong 3f_{q}^{2}(N-Z).
\label{Eqt. ISR}
\end{equation}

The reduced transition probabilities for GT transitions from the
QRPA ground state to one-phonon states in the daughter nucleus were
obtained as
\begin{equation}
B_{GT}^{\pm}(\omega) = |\langle \omega, \mu
\|t_{\pm}\sigma_{\mu}\|QRPA\rangle|^{2}, \label{Eqt. 7}
\end{equation}
where the symbols have their usual meaning. $\omega$ represents
daughter excitation energies. $\mu$ can only take three values (-1,
0, 1) and represents the third component of the angular momentum.
The charge-changing transition strengths were calculated as in
Eq.~(\ref{Eqt. 7}). For details of calculation of nuclear matrix
elements we refer to \cite{Nabi04}.

For odd-A nuclei, there exist two different types of transitions:
(a) phonon transitions with the odd particle acting only as a
spectator and (b) transitions of the odd particle itself. For case
(b) phonon correlations were introduced to one-quasiparticle states
in first-order perturbation. For further details, we refer to
\citep{Hir93}.

\subsection{Neutrino-Antineutrino Energy Loss Rates}

As discussed earlier the neutrino and antineutrino energy loss rates
can occur through four different weak-interaction mediated channels:
electron and positron emissions, and, continuum electron and
positron captures. It is assumed that the neutrinos and
antineutrinos produced as a result of these reactions are
transparent to the stellar matter during the presupernova
evolutionary phases and contributes effectively in cooling the
system. The neutrino and antineutrino energy loss rates were
calculated using the relation

\begin{equation}
\lambda ^{^{\nu(\bar{\nu})}} _{ij} = \left(\frac{ln 2}{D} \right)
[f_{ij}^{\nu} (T, \rho, E_{f})][B(F)_{ij} + (g_{A}/ g_{V})^{2}
B(GT)_{ij}]. \label{wi}
\end{equation}
The value of D was taken to be 6295s \citep{Yos88}. $B_{ij}'s$ are
the sum of reduced transition probabilities of the Fermi B(F) and
Gamow-Teller (GT) transitions B(GT). The effective ratio of axial
and vector coupling constants, $(g_{A}/g_{V})$ was taken to be
-1.254 \citep{Rod06}. The $f_{ij}^{\nu}$ are the phase space
integrals and are functions of stellar temperature ($T$), density
($\rho$) and Fermi energy ($E_{f}$) of the electrons. They are
explicitly given by
\begin{equation}
f_{ij}^{\nu} \, =\, \int _{1 }^{w_{m}}w\sqrt{w^{2} -1} (w_{m} \,
 -\, w)^{3} F(\pm Z,w)(1- G_{\mp}) dw,
\label{phdecay}
\end{equation}
and by
\begin{equation}
f_{ij}^{\nu} \, =\, \int _{w_{l} }^{\infty }w\sqrt{w^{2} -1} (w_{m}
\,
 +\, w)^{3} F(\pm Z,w)G_{\mp} dw.
\label{phcapture}
\end{equation}
In Eqs. ~(\ref{phdecay}) and ~(\ref{phcapture})  $w$ is the total
energy of the electron including its rest mass, $w_{l}$ is the total
capture threshold energy (rest+kinetic) for positron (or electron)
capture. F($ \pm$ Z,w) are the Fermi functions and were calculated
according to the procedure adopted by  \citep{Gov71}. G$_{\pm}$ is
the Fermi-Dirac distribution function for positrons (electrons).
\begin{equation}
G_{+} =\left[\exp \left(\frac{E+2+E_{f} }{kT}\right)+1\right]^{-1},
\label{Gp}
\end{equation}
\begin{equation}
 G_{-} =\left[\exp \left(\frac{E-E_{f} }{kT}
 \right)+1\right]^{-1},
\label{Gm}
\end{equation}
here $E$ is the kinetic energy of the electrons and $k$ is the
Boltzmann constant.

For the decay (capture) channel Eq. ~(\ref{phdecay}) (Eq.
~(\ref{phcapture})) was used for the calculation of phase space
integrals. Upper (lower) signs were used for the case of electron
(positron) emissions in Eq.~(\ref{phdecay}). Similarly upper (lower)
signs were used for the case of continuum electron (positron)
captures in Eq. ~(\ref{phcapture}). Details of the calculation of
reduced transition probabilities can be found in Ref.
\citep{Nabi04}. Construction of parent and daughter excited states
and calculation of transition amplitudes between these states can be
seen in Ref. \citep{Nabi99a}.

The total neutrino energy loss rate per unit time per nucleus is
given by
\begin{equation}
\lambda^{\nu} =\sum _{ij}P_{i} \lambda _{ij}^{\nu}, \label{nurate}
\end{equation}
where $\lambda_{ij}^{\nu}$ is the sum of the electron capture and
positron decay rates for the transition $i \rightarrow j$ and
$P_{i}$ is the probability of occupation of parent excited states
which follows the normal Boltzmann distribution.

On the other hand the total antineutrino energy loss rate per unit
time per nucleus is given by
\begin{equation}
\lambda^{\bar{\nu}} =\sum _{ij}P_{i} \lambda _{ij}^{\bar{\nu}},
\label{nubarrate}
\end{equation}
where $\lambda_{ij}^{\bar{\nu}}$ is the sum of the positron capture
and electron decay rates for the transition $i \rightarrow j$.

\section{Results and Discussions}
\label{sec:results}

\begin{figure}
\begin{center}
\includegraphics[width=0.7\textwidth]{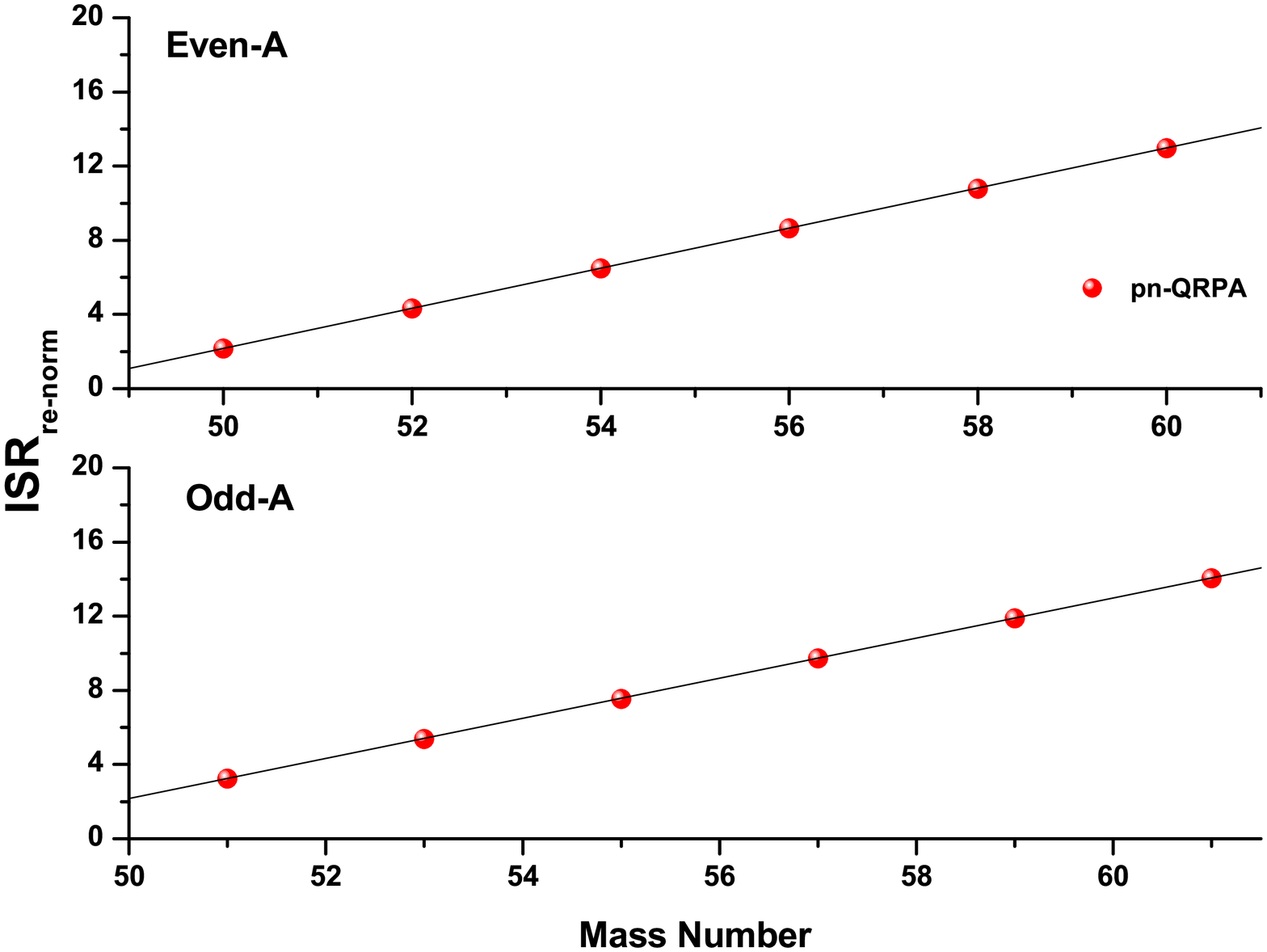}
\end{center}
\caption{Calculated re-normalized Ikeda Sum Rule. The straight line
is the theoretical prediction of the sum rule and is shown just to
guide the eye.} \label{RISR}
\end{figure}

The $\beta$-decay and capture rates are exponentially sensitive to
the location of GT$_{+}$ resonance \citep{Auf96} which in turn
translates to the placement of GT centroid in daughter. The
statistical data for the calculated GT strength distributions for
isotopes of chromium ($^{50-60}$Cr) is presented in Table~\ref{ta1}.
Here we show the calculated GT strengths (in arbitrary units),
centroids (in MeV) and widths (in MeV) along both $\beta$-decay and
electron capture direction for isotopes of chromium. The fulfillment
of Ikeda Sum Rule (Eq.~(\ref{Eqt. ISR})) is one of the key factors
to check for the consistency of any theoretical calculation of GT
strength function. Fig.~\ref{RISR} shows the excellent comparison of
our calculated re-normalized Ikeda Sum Rule with the theoretical
prediction.

We compare our calculated total GT strengths with other theoretical
calculations and measurements wherever possible in Table~\ref{ta2}.
References for previous theoretical calculations and experimental
data is provided in caption of Table~\ref{ta2}. It is to be noted
that the pn-QRPA calculated strengths are relatively smaller than
those calculated by shell model results. The total $B(GT)_{+}$
strength for $^{56}$Cr calculated by SMMC(KB3) model is 1.5$\pm$0.21
\citep{Lan95} (not shown in Table~\ref{ta2}). The corresponding
strength calculated by our model is 1.31 and is in decent comparison
with the shell model result.

\begin{figure}
\begin{center}
\includegraphics[width=0.7\textwidth]{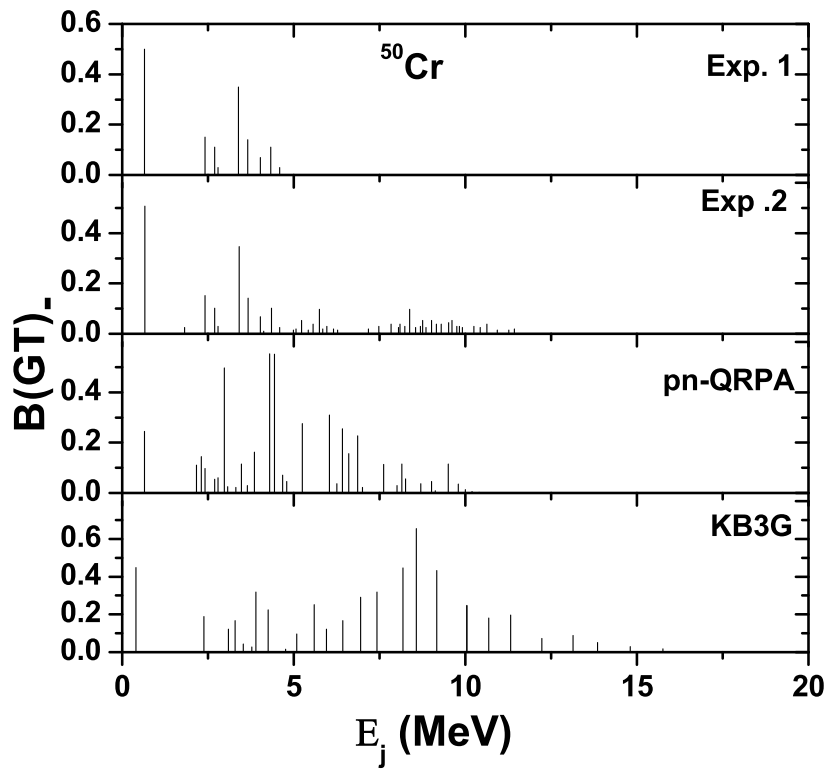}
\end{center}
\caption{Comparison of calculated B(GT)$_{-}$ strength distributions
in $^{50}$Cr with measurements and other theoretical model. Exp. 1
shows measured values by \citep{Fuj11}, Exp. 2 by \citep{Ada07}
while KB3G shows shell model calculation by \citep{Pet07}. $E_{j}$
represents excitation energy in $^{50}$Mn in units of MeV.}
\label{50cr}
\end{figure}

Fig.~\ref{50cr} shows our calculated GT strength distribution in the
$\beta$-decay direction for $^{50}$Cr. Shown also are the two
measured GT distributions for $^{50}$Cr. Exp. 1 shows the
measurement result of ($^{3}$He, $t$) experiment up to 5 MeV by
\cite{Fuj11}. The high resolution $^{50}$Cr($^{3}$He, $t$)$^{50}$Mn
measurement at an incident energy of 140 MeV/nucleon and at 0$^{0}$,
for a precise study of GT transitions up to 12 MeV in daughter
performed by \cite{Ada07}, is shown as  Exp. 2 in Fig.~\ref{50cr}.
We further compare our calculated GT strength distribution with the
previous shell model calculation of \cite{Pet07} using the KB3G
interaction. Fragmentation of GT $1^+$ strength exists in all cases.
It can be seen that the pn-QRPA calculates low-lying transitions in
daughter of bigger magnitude than the shell model results resulting
in placement of centroid at a much lower energy in daughter. The
pn-QRPA calculated strength distribution is in  good agreement with
Exp. 2 data.

\begin{figure}
\begin{center}
\includegraphics[width=0.7\textwidth]{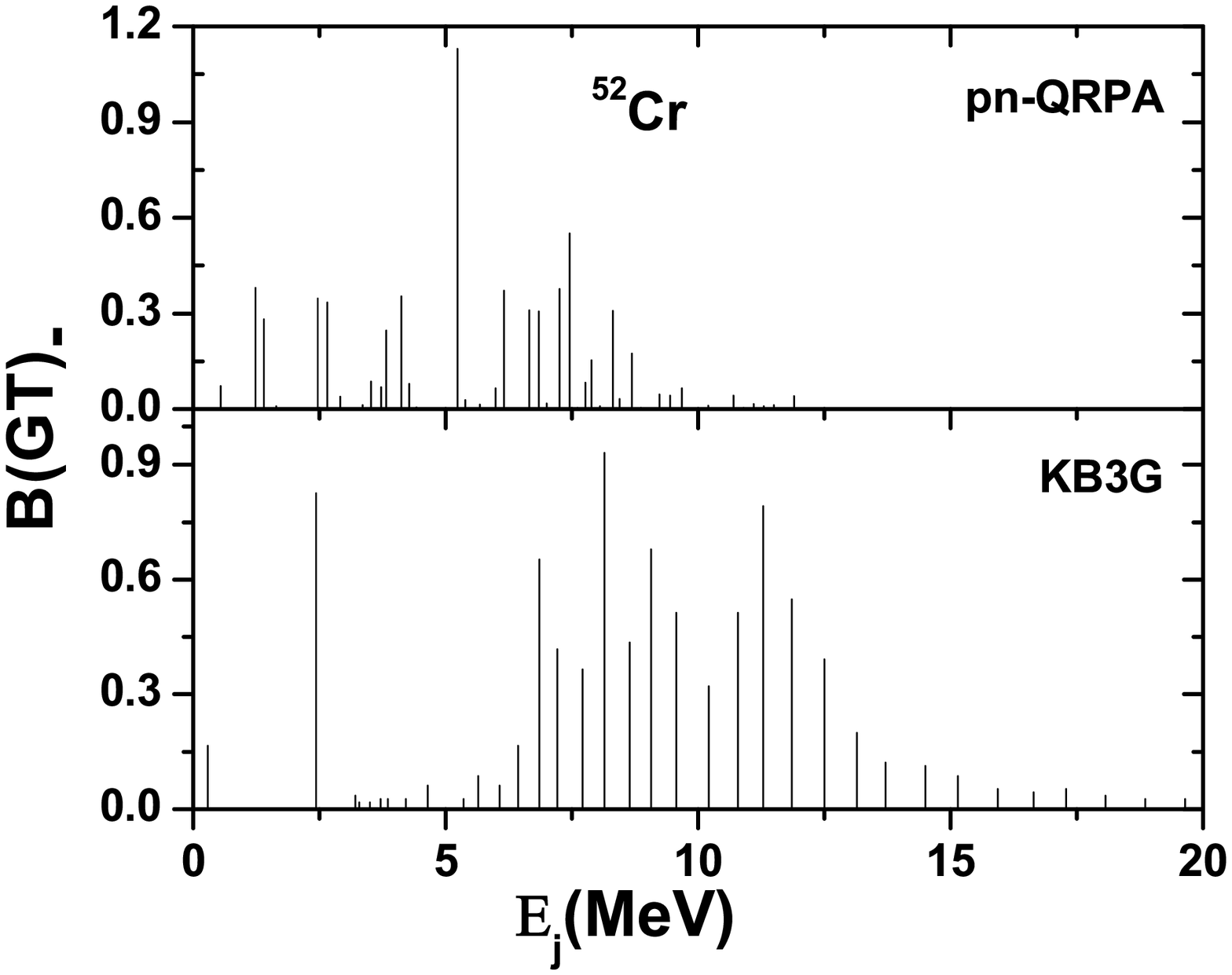}
\end{center}
\caption{Comparison of calculated B(GT)$_{-}$ strength distributions
in $^{52}$Cr with shell model calculation \citep{Pet07}. $E_{j}$
represents excitation energy in $^{52}$Mn in units of MeV.}
\label{52cr}
\end{figure}

\citep{Pet07} also performed a large scale shell model calculation
of GT$_{-}$ in $^{52}$Cr using the KB3G interaction. We compare
their results with our pn-QRPA calculation in Fig.~\ref{52cr}.  In
shell model calculation the GT states are mainly concentrated
between 5--15 MeV in daughter. The pn-QRPA places the energy
centroid at low excitation energy of 5.41 MeV in $^{52}$Mn.

\begin{figure}
\begin{center}
\includegraphics[width=0.7\textwidth]{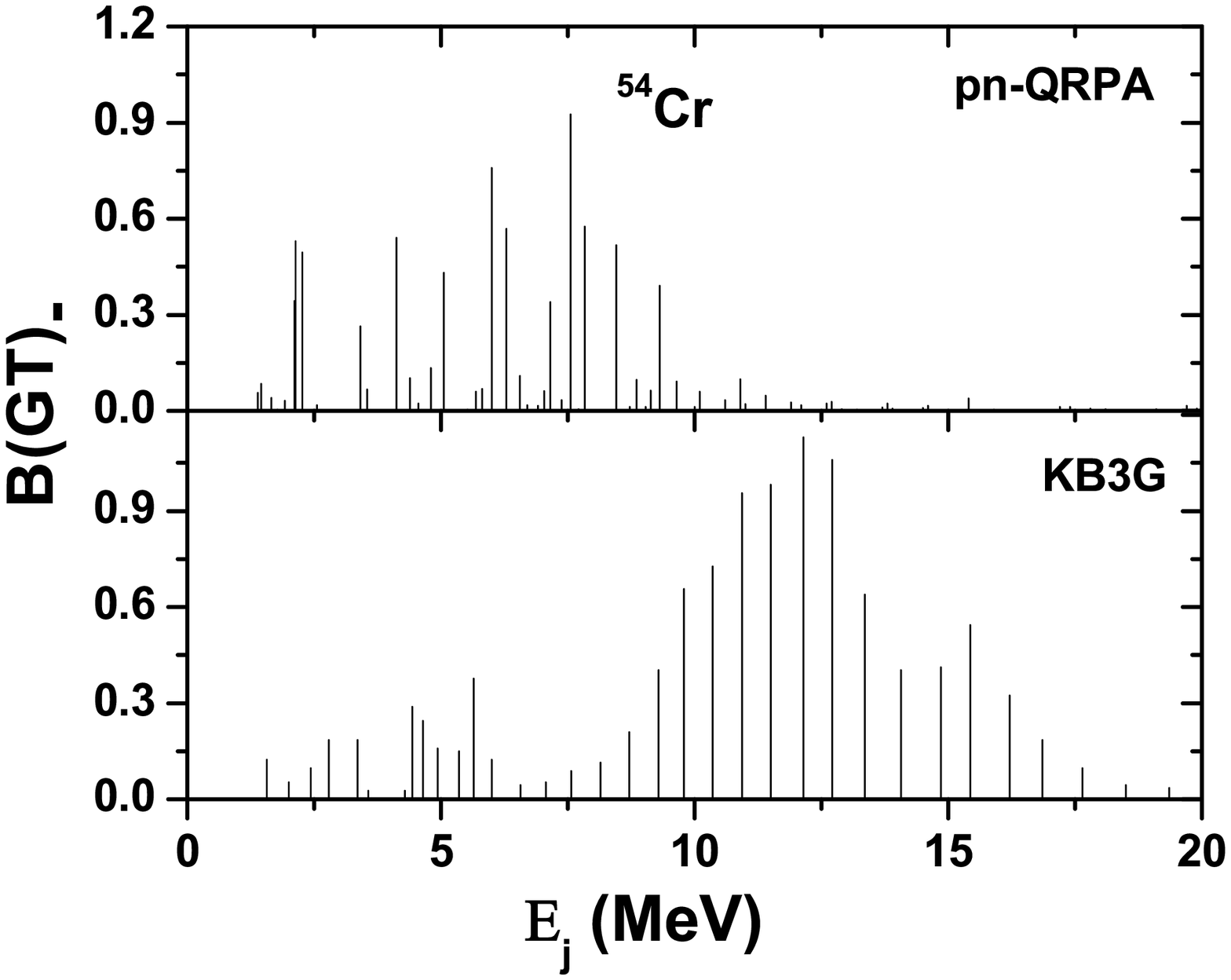}
\end{center}
\caption{Comparison of calculated B(GT)$_{-}$ strength distributions
in $^{54}$Cr with shell model calculation \citep{Pet07}. $E_{j}$
represents excitation energy in $^{54}$Mn in units of MeV.}
\label{54cr}
\end{figure}

A similar comparison of our calculated GT$_{-}$ in $^{54}$Cr with
that of \cite{Pet07} is shown in Fig.~\ref{54cr}.  Bulk of GT
strength in $1^+$ states have been concentrated in different energy
ranges in both models. They are placed at energy intervals of
2.5--10 MeV in pn-QRPA model and 8--16 MeV in shell model
calculation.

The calculated neutrino and antineutrino loss rates due to 11
isotopes of chromium ($^{50-60}$Cr) for selected densities and
temperatures in stellar matter are presented in
Tables~\ref{ta3}-\ref{ta5}. The first column of the tables gives
$\log \rho \text{Y}_{e}$ in units of g\;cm$^{-3}$, where $\rho$ is
the baryon density and Y$_{e}$ is the ratio of the lepton number to
the baryon number. Stellar temperatures (T$_{9}$) are given in units
of $10^{9}$\;K. $\lambda_{\nu}$ ($\lambda_{\bar{\nu}}$) are the
total neutrino (antineutrino) energy loss rates  as a result of
$\beta^{+}$ decay and electron capture ($\beta^{-}$ decay and
positron capture) in units of MeV\;s$^{-1}$. All calculated rates
are tabulated in logarithmic (to base 10) scale. In the tables, -100
means that the rate is smaller than 10$^{-100}$ MeV\;s$^{-1}$. It
can be seen from Table~\ref{ta3} that at low stellar temperatures
the neutrino energy loss rates due to $^{50,51}$Cr dominate by order
of magnitudes. As temperature soars to $T_{9}[\text{K}] \sim 30$,
the antineutrino energy loss rates try to catch up with the neutrino
energy loss rates. For $^{52}$Cr, energy losses by neutrino and
antineutrino have comparable rates in density range $\rho$ = 10$^{2
- 5}$ g\;cm$^{-3}$, while for $^{53}$Cr, at same density range, the
antineutrino energy loss rates dominate. At high stellar densities
the neutrino energy loss rates  due to $^{52,53}$Cr are orders of
magnitude bigger.  The calculated energy losses due to weak rates on
$^{54-57}$Cr and $^{58-60}$Cr are given in Tables~\ref{ta4}
and~\ref{ta5}, respectively. It can be seen from
Tables~\ref{ta3}-\ref{ta5} that at low densities and temperatures
the antineutrino energy loss rates due to $^{53-60}$Cr dominate by
order of magnitudes and hence more important for the collapse
simulators. As T$_{9} [K] \sim 30$, the neutrino energy loss rates
try to catch up with the antineutrino energy loss rates. At high
stellar densities the story reverses with neutrino energy loss rates
assuming the role of the dominant partner. At low densities the
antineutrino energy loss rates have a dominant contribution from the
positron captures. As temperature rises or density lowers (the
degeneracy parameter is negative for positrons), more and more
high-energy positrons are created leading in turn to higher positron
capture rates and consequently higher antineutrino energy loss
rates. The complete electronic version (ASCII files) of these rates
may be requested from the authors.

Our calculation of neutrino and antineutrino energy loss rates due
to weak interactions on chromium isotopes was also compared with
previous calculations performed by \citep{Fuller} (FFN) and those
performed using the large-scale shell model (LSSM) by \cite{Lan00}.
The FFN rates had been used in many simulation codes (e.g., KEPLER
stellar evolution code) while LSSM rates were employed in recent
simulation of presupernova evolution of massive stars in the mass
range 11-40\;M$_\odot$ \cite{Heg01}. Here we compare our calculation
of (anti)neutrino energy loss rates for all isotopes of chromium
which were found to be astrophysically important as per simulation
results of \citep{Aufderheide94, Heg01}.

\begin{center}
\begin{figure}[tb]
  \begin{center}
  \includegraphics[width=0.7\textwidth]{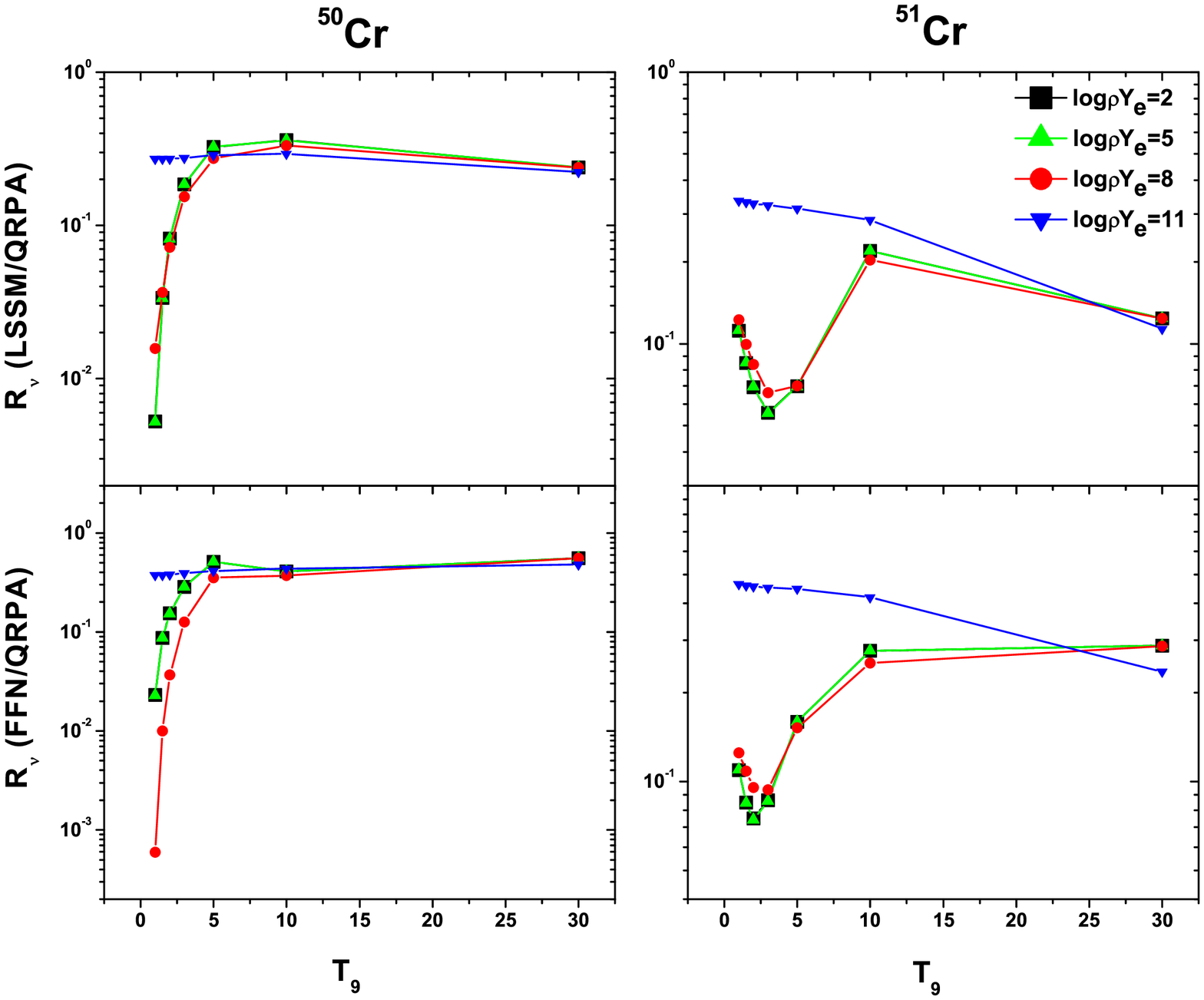}
  \end{center}
\caption{\small The comparison of pn-QRPA neutrino energy loss rates
due to $^{50}$Cr (left) and $^{51}$Cr (right) with the previous
calculations performed by LSSM (upper panel) and those performed by
FFN (lower panel) as function of stellar temperatures for different
selected densities. $\log\rho \text{Y}_{e}$ gives the $\log$ to base
10 of stellar density in units of g\;cm$^{-3}$. T$_{9}$ gives the
stellar temperature in units of $10^9$\;K.} \label{50-51}
\end{figure}
\end{center}

Figure~\ref{50-51} shows the comparison of neutrino energy loss
rates due to weak rate interactions on $^{50}$Cr (left column) and
$^{51}$Cr (right column) with the FFN and LSSM calculations. The
upper panel displays the ratio of the LSSM rates to the calculated
rates, R$_{\nu}$(LSSM/QRPA), and the lower panel shows a similar
comparison with the FFN calculation, R$_{\nu}$(FFN/QRPA). All graphs
are drawn at four selected densities ($\rho
\text{Y}_{e}$[g\;cm$^{-3}$] = $10^{2}, 10^{5}, 10^{8}$ and
$10^{11}$). These values correspond to low, medium-low, medium-high
and high stellar densities, respectively. The calculated ratios are
shown as a function of stellar temperatures ranging from T$_{9}
[\text{K}]$ = 1 to 30. Our calculated neutrino energy loss rates due
to  $^{50}$Cr is more than two orders of magnitude (factor 43)
bigger than the rates calculated by LSSM (FFN) at T$_{9} [\text{K}]$
= 1 at low and medium-low densities. As stellar temperature soars to
T$_{9} [\text{K}]$ = 30, our rates are still factor 5 (2) bigger
than LSSM (FFN) rates. At high stellar densities and  temperatures
the mutual comparison with previous calculations improves to within
a factor two.  The primary reason for our enhanced neutrino energy
loss rates may be traced back to the calculation of our ground-state
GT strength distributions. Our calculated GT strength distribution
centroids reside at much lower energy in daughter than shell model
calculation (see Figs.~\ref{50cr}-~\ref{54cr}). To a lesser extent,
the difference in calculated rates may also be attributed to
calculation of excited state GT strength distributions in the two
models. The LSSM employed the so-called Brink's hypothesis in the
electron capture direction and back-resonances in the $\beta$-decay
direction to approximate the contributions from high-lying excited
state GT strength distributions in their calculation of weak rates.
Brink's hypothesis states that GT strength distribution on excited
states is \textit{identical} to that from ground state, shifted
\textit{only} by the excitation energy of the state. GT back
resonances are the states reached by the strong GT transitions in
the inverse process (electron capture) built on ground and excited
states. On the other hand the pn-QRPA model performs a microscopic
calculation of the GT strength distributions for \textit{all} parent
excited states and provides a fairly reliable estimate of the total
stellar rates. For the case of $^{51}$Cr (right column) the pn-QRPA
neutrino energy loss rates are again bigger for reasons mentioned
earlier. At high densities the comparison improves with LSSM and FFN
calculations. However our rates are still bigger by factor of 3-9.
Simulators should take note of our enhanced neutrino energy loss
rates at low stellar temperatures and densities characteristic of
the hydrostatic phases of stellar evolution which may affect the
temperature and the corresponding lepton-to baryon ratio which
becomes very important going into stellar collapse.

\begin{center}
\begin{figure}[tb]
  \begin{center}
  \includegraphics[width=0.7\textwidth]{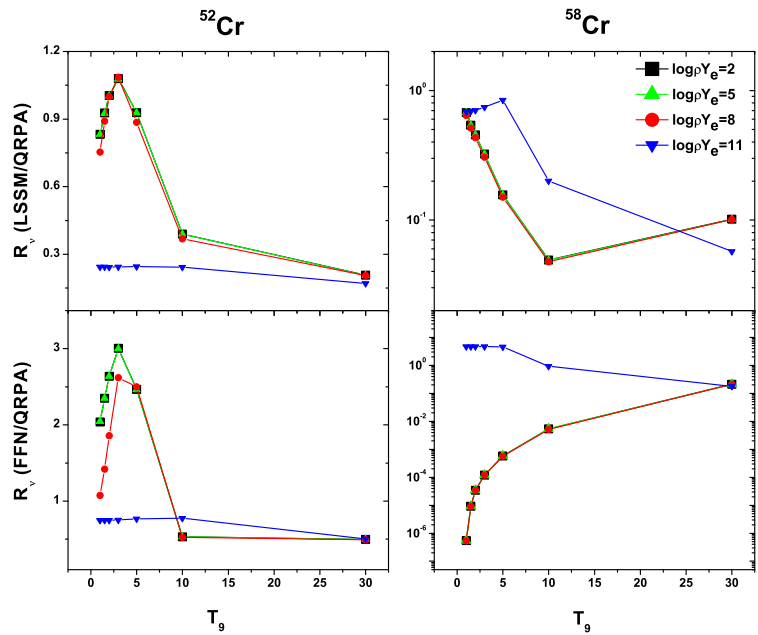}
  \end{center}
\caption{\small Same as Fig.~\ref{50-51} but for neutrino energy
loss rates due to $^{52}$Cr (left) and $^{58}$Cr (right)}
\label{52-58}
\end{figure}
\end{center}

The left panels of Figure~\ref{52-58} shows that our calculation of
neutrino energy loss rates due to $^{52}$Cr agree  with the previous
calculations to within a factor 5. For the case of $^{58}$Cr (right
panels) we note that FFN rates are up to 7 orders of magnitude
smaller than our rates and surpass our calculated rates only at high
stellar densities. Our results are in better comparison with the
LSSM calculation. Unmeasured matrix elements for allowed transitions
were assigned an average value of $log ft = $5 in FFN calculations.
On the other hand these transitions were calculated in a microscopic
fashion using the pn-QRPA theory (and LSSM) and depict a more
realistic picture of the events taking place in stellar environment.

\begin{center}
\begin{figure}[tb]
  \begin{center}
  \includegraphics[width=0.7\textwidth]{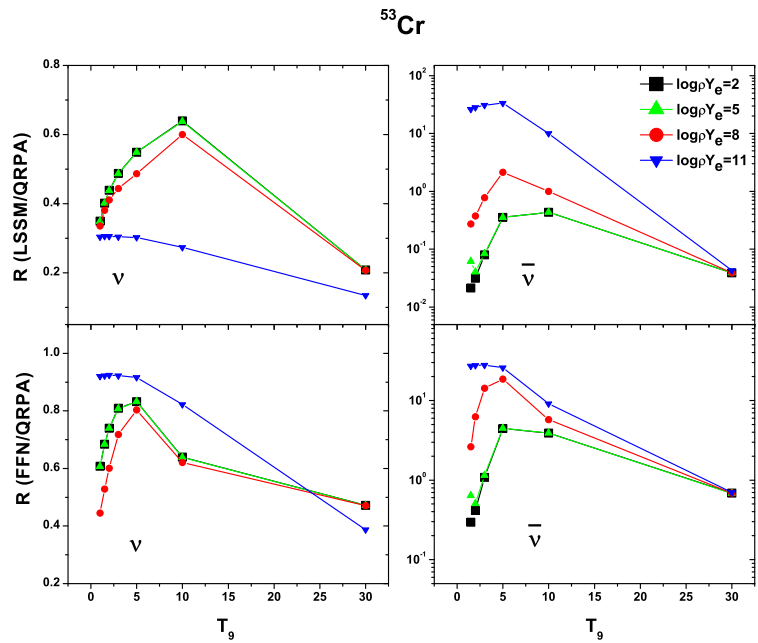}
  \end{center}
\caption{\small The comparison of pn-QRPA neutrino and antineutrino
energy loss rates due to $^{53}$Cr with the previous calculations
performed by LSSM (upper panel) and those performed by FFN (lower
panel) as function of stellar temperatures for different selected
densities. $\log\rho \text{Y}_{e}$ gives the $\log$ to base 10 of
stellar density in units of g\;cm$^{-3}$. T$_{9}$ gives the stellar
temperature in units of $10^9$\;K.} \label{53}
\end{figure}
\end{center}

\begin{center}
\begin{figure}[tb]
  \begin{center}
  \includegraphics[width=0.7\textwidth]{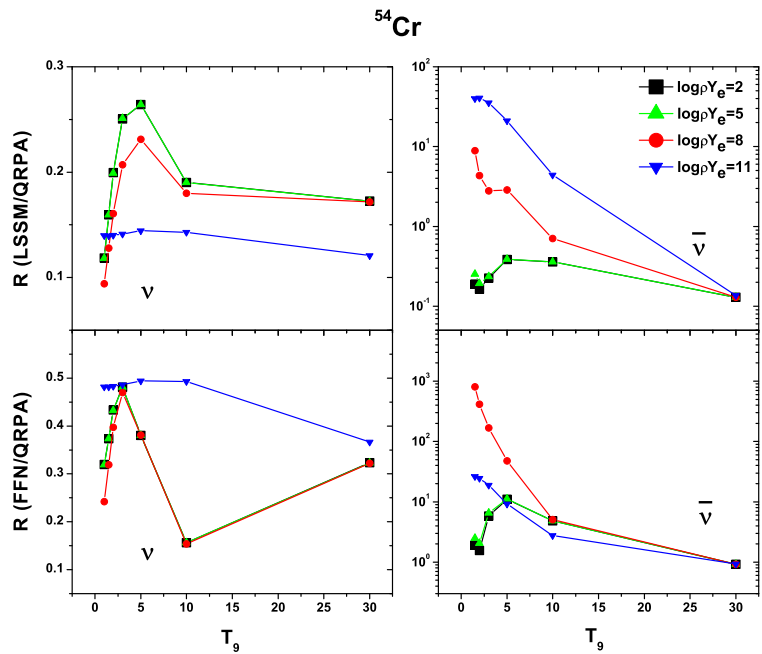}
  \end{center}
\caption{\small Same as Fig.~\ref{53} but due to $^{54}$Cr.}
\label{54}
\end{figure}
\end{center}

\begin{center}
\begin{figure}[tb]
  \begin{center}
  \includegraphics[width=0.7\textwidth]{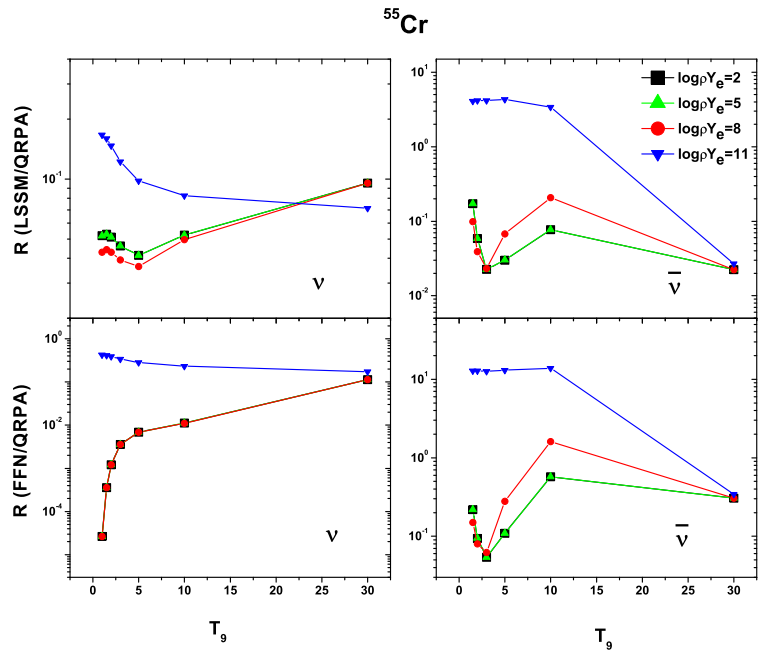}
  \end{center}
\caption{\small Same as Fig.~\ref{53} but due to $^{55}$Cr.}
\label{55}
\end{figure}
\end{center}

\begin{center}
\begin{figure}[tb]
  \begin{center}
  \includegraphics[width=0.7\textwidth]{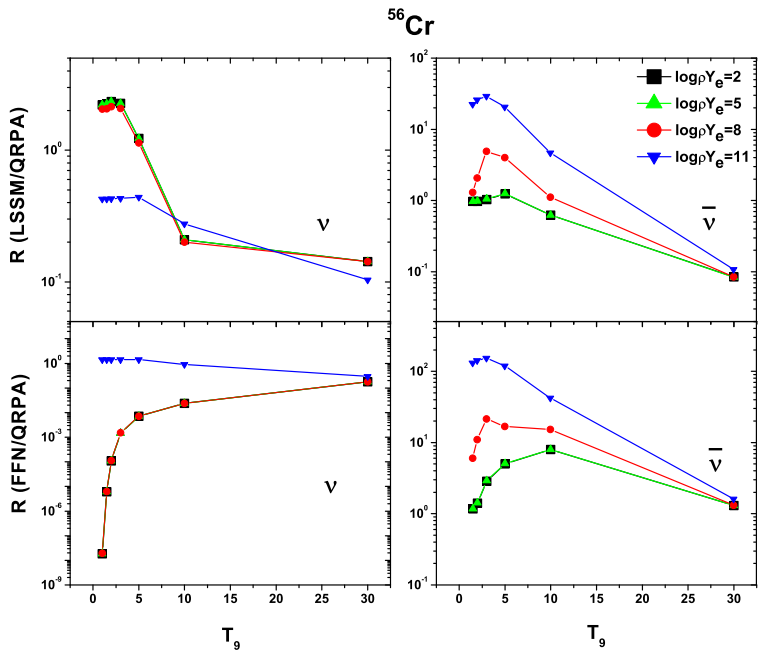}
  \end{center}
\caption{\small Same as Fig.~\ref{53} but due to $^{56}$Cr.}
\label{56}
\end{figure}
\end{center}

\begin{center}
\begin{figure}[tb]
  \begin{center}
  \includegraphics[width=0.7\textwidth]{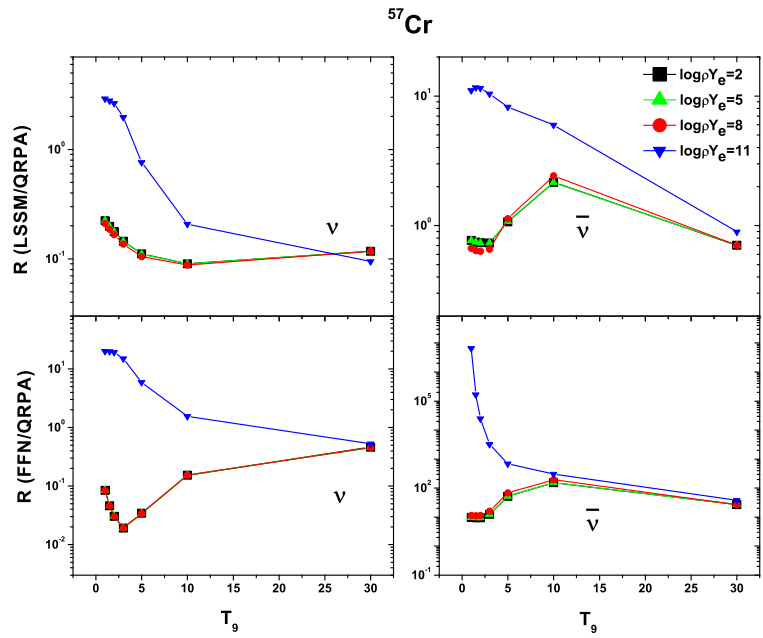}
  \end{center}
\caption{\small Same as Fig.~\ref{53} but due to $^{57}$Cr.}
\label{57}
\end{figure}
\end{center}

Figures~\ref{53}-\ref{57} show the simultaneous comparison of
neutrino (left panels) and antineutrino energy loss rates (right
panels) with previous calculations for $^{53}$Cr to $^{57}$Cr,
respectively. Figure~\ref{53} shows that for the case of $^{53}$Cr
the three neutrino energy loss rate calculations are in decent
comparison whereas orders of magnitude differences are seen in the
comparison of antineutrino energy loss rates. Our calculated
antineutrino energy loss rates are in good comparison with FFN for
low and medium-low  density regions. At higher densities FFN rates
are bigger by more than one order of magnitude. There are two main
reasons for this enhancement of FFN rates. Firstly, FFN placed the
centroid of the GT strength at too low excitation energies in their
compilation of weak rates for odd-A nuclei \cite{Lan98}. Secondly,
FFN threshold parent excitation energies were not constrained and
extended well beyond the particle decay channel. At high
temperatures contributions from these high excitation energies begin
to show their cumulative effect.  Our antineutrino energy loss rates
are more than an order of magnitude bigger than LSSM at low
densities and temperatures. The reason is the calculation of more GT
strength at lower energies in daughter in our model as discussed
earlier. LSSM rates get bigger as stellar density increases.

\begin{center}
\begin{figure}[tb]
  \begin{center}
  \includegraphics[width=0.7\textwidth]{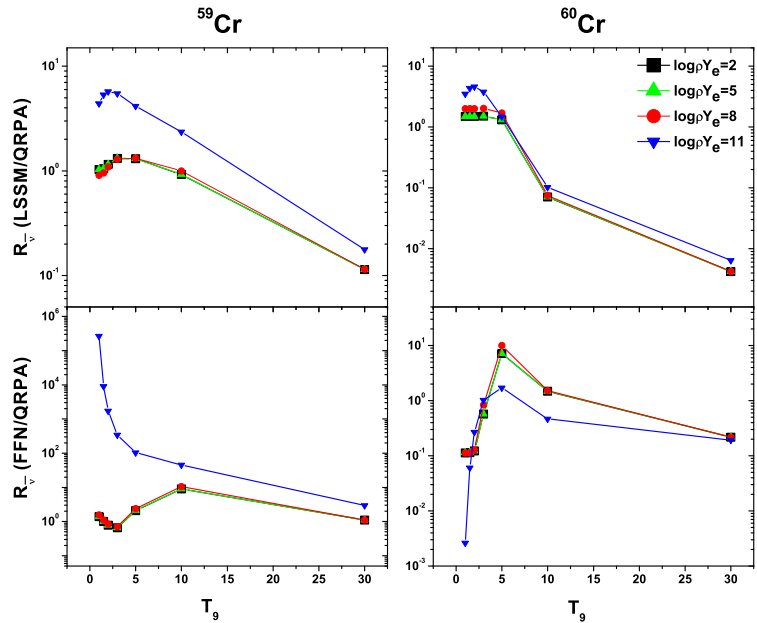}
  \end{center}
\caption{\small Same as Fig.~\ref{50-51} but for antineutrino energy
loss rates due to $^{59}$Cr (left) and $^{60}$Cr (right)}
\label{59-60}
\end{figure}
\end{center}

A similar comparison is seen in Figure~\ref{54} for the case of
neutrino energy loss rate due to $^{54}$Cr.  Our calculated
antineutrino energy loss rates are bigger than LSSM at low densities
and temperatures. At high densities LSSM rates get bigger except at
high temperatures (for reasons already stated). The FFN rates are
bigger. It is further to be noted that FFN neglected the quenching
of the total GT strength in their rate calculation. The pn-QRPA
calculated neutrino energy loss rates due to $^{55}$Cr
(Figure~\ref{55}) are orders of magnitude bigger than FFN. The
approximations used by FFN in calculation of  nuclear matrix
elements were not good and resulted in very small neutrino cooling
rates as compared with the microscopic calculations performed by us
and LSSM. Our calculated neutrino energy loss rates due to $^{55}$Cr
are also bigger than LSSM results. Only at high density does the
comparison improves. Our calculated antineutrino energy loss rates
due to $^{55}$Cr are bigger at low, medium-low and medium-high
densities. At high density, the FFN and LSSM rates get factor 5-10
bigger. Our calculated neutrino energy loss rates for the case of
$^{56}$Cr  are orders of magnitude bigger than FFN
(Figure~\ref{56}). The situation is very much similar to the
comparison seen in bottom-left panel of Figure~\ref{55}. The reason
for this large discrepancy was stated earlier. At high density the
two results compare well. The antineutrino energy loss rates are in
better comparison. For the case of $^{57}$Cr (Figure~\ref{57}) our
calculated neutrino energy loss rates are generally bigger  except
at high density. The antineutrino energy loss rates of FFN are
bigger by orders of magnitude at high density. The antineutrino
energy loss rate comparison is fair for the case of $^{59,60}$Cr
(Figure~\ref{59-60}). At high density FFN rates are too big for
$^{59}$Cr whereas a decent comparison is seen for the case of
$^{60}$Cr. It is to be noted that both pn-QRPA theory and LSSM
calculates the ground-state GT distributions microscopically. For
higher lying excited states, pn-QRPA model again calculates the GT
strength distributions in a microscopic fashion whereas Brink's
hypothesis and back resonances are employed in LSSM and FFN
calculations. Accordingly, whenever ground state rates command the
total rate, the two calculations are found to be in excellent
agreement. For cases where excited state partial rates influence the
total rate, differences are seen between the two calculations.

\section{Conclusions}
\label{sec:conclusions}

For stellar densities less than $\rho \text{Y}_{e}$[g\;cm$^{-3}$] =
$10^{11}$, the non-thermal (anti)neutrinos produced as a result of
weak-interaction rates are transparent to stellar matter and cools
the stellar core as a result of energy transfer. This process also
reduces the entropy of the core material. In this paper we
concentrated on astrophysically important isotopes of chromium and
calculated  their GT transitions using the microscopic pn-QRPA
theory. The calculated GT strength distributions satisfied the
model-independent Ikeda Sum Rule and were found in decent comparison
with measured data wherever available. We also calculated the
centroids and widths of the calculated GT strength distributions for
11 isotopes of chromium.

Later we performed calculation of (anti)neutrino energy loss rates
due to these isotopes of chromium in stellar matter. The neutrino
and antineutrino energy loss rates were calculated on a detailed
density-temperature grid point and the ASCII files of the rates can
be requested from the authors. The rates were also compared with the
previous calculations (LSSM and FFN). FFN and LSSM calculations used
approximations like Brink's hypothesis, back-resonances not used in
our calculation. FFN calculation suffered with problems like
placement of centroids of GT strengths, microscopic calculation of
nuclear matrix elements, quenching of GT strength. On the other hand
the LSSM calculation possessed the convergence problem
(Lanczos-based as pointed by \cite{Pru03}). Our pn-QRPA model did
not suffer from these issues and we believe our calculated rates
provide a fair and realistic picture of energy transfer from stellar
cores via (anti)neutrino carriers.

We will urge simulators to test run our reported weak interaction
rates presented here to check for some interesting outcome. We are
currently in a phase of extending the present work for other nuclide
of astrophysical importance and hope to report on the outcome of
these calculations in near future.

\acknowledgments J.-U. Nabi would like to acknowledge the support of
the Higher Education Commission Pakistan through the HEC Project No.
20-3099.

\clearpage \onecolumn

\clearpage \onecolumn

\clearpage

\begin{table}\tiny
  \begin{center}
  \caption{Total strength, centroid and width of calculated GT strength distributions (for $\beta$-decay and electron capture directions) for  $^{50-60}$Cr.}\label{ta1}
    \begin{tabular}{ccccccc}
    Nuclei & $\sum B(GT)_{-}$ & $\sum B(GT)_{+}$ & $\bar{E}_{-}$ (MeV) &
    $\bar{E}_{+}$ (MeV) & Width$_{-}$ (MeV)& Width$_{+}$ (MeV)\\
    \hline
    $^{50}$Cr  & 4.65  & 2.49  & 4.81  & 4.16  & 2.12  & 2.3   \\
    $^{51}$Cr & 5.13  & 1.87  & 7.72  & 7.96  & 3.06  & 2.41   \\
  $^{52}$Cr  & 6.55  & 2.21  & 5.41  & 3.27  & 2.44  & 1.92   \\
   $^{53}$Cr  & 5.91  & 0.51  & 8.81  & 6.21  & 2.79  & 2.71  \\
    $^{54}$Cr  & 8.45  & 1.95  & 6.19  & 2.88  & 2.97  & 3.32  \\
    $^{55}$Cr  & 7.94  & 0.39  & 9.57  & 4.06  & 3.05  & 3.47   \\
   $^{56}$Cr  & 9.95  & 1.31  & 6.44  & 1.77  & 2.59  & 2.14   \\
   $^{57}$Cr & 9.98  & 0.25  & 9.62  & 5.21  & 2.98  & 2.84   \\
    $^{58}$Cr  & 11.6  & 0.82  & 6.85  & 1.57  & 2.83  & 2.49   \\
   $^{59}$Cr  & 12.1  & 0.24  & 6.86  & 1.26  & 4.79  & 2.24  \\
    $^{60}$Cr  & 13.4  & 0.39  & 7.74  & 3.03  & 3.3   & 4.99   \\
    \end{tabular}
\end{center}
  \end{table}

\begin{table}
\tiny \caption{Comparison of calculated total $B(GT)$ values in
$^{50,52,53,54}$Cr with measurement and other theoretical models
(LSSM (KB3G) $\rightarrow$ \cite{Pet07}, SMMC (KB3) $\rightarrow$
\cite{Lan95}, Shell Model $\rightarrow$ \cite{Nak96} and Shell Model
(KB3) $\rightarrow$ \cite{Cau95}). Experimental data was taken from
\cite{Ada07}. }\label{ta2}
\begin{tabular}{c|cc|cc|cc|cc}
&$^{50}$Cr&&$^{52}$Cr&&$^{53}$Cr&&$^{54}$Cr\\
&$\sum B(GT)_{-}$&$\sum B(GT)_{+}$&$\sum B(GT)_{-}$&$\sum B(GT)_{+}$&$\sum B(GT)_{-}$&$\sum B(GT)_{+}$ &$\sum B(GT)_{-}$&$\sum B(GT)_{+}$ \\
\hline
pn-QRPA &4.65&2.49&6.55&2.22&5.90&0.51&8.44&1.96\\
LSSM (KB3G) &5.20& - &8.85&-&-&-&11.13&- \\
SMMC (KB3)&-&$3.51\pm 0.27$&-&3.51$\pm$0.19&-&-&-&2.21$\pm$0.22 \\
Shell Model &-&-&17.4&4.3&20.1&3.8&22.4&2.9 \\
Shell Model (KB3)&-&3.57&-&-&-&-&-&- \\
Exp.  &2.69&-&-&-&-&-&-&- \\
\hline
\end{tabular}
\end{table}

\begin{table}[pt]
\caption{Neutrino and antineutrino energy loss rates due to
$^{50,51,52,53}$Cr for selected densities and temperatures in
stellar matter. $\log\rho \text{Y}_{e}$ has units of g\;cm$^{-3}$,
where $\rho$ is the baryon density and Y$_{e}$ is the ratio of the
lepton number to the baryon number. Temperatures (T$_{9}$) are given
in units of $10^{9}$\;K. $\lambda_{\nu}$ ($\lambda_{\bar{\nu}}$) are
the total neutrino (antineutrino) energy loss rates  as a result of
$\beta^{+}$ decay and electron capture ($\beta^{-}$ decay and
positron capture) in units of MeV\;s$^{-1}$. All calculated rates
are tabulated in logarithmic (to base 10) scale. In the table, -100
means that the rate is smaller than 10$^{-100}$ MeV\;s$^{-1}$.}
\label{ta3} {\scriptsize\begin{tabular}{|cc|cccccccc|} $\log\rho
\text{Y}_{e}$ & T$_{9}$ & \multicolumn{2}{c|}{$^{50}$Cr}&
\multicolumn{2}{c|}{$^{51}$Cr} & \multicolumn{2}{c|}{$^{52}$Cr}
& \multicolumn{2}{c|}{$^{53}$Cr} \\
\cline{3-10} & &  \multicolumn{1}{c}{$\lambda_{\nu}$} &
\multicolumn{1}{|c|}{$\lambda_{\bar{\nu}}$} &
\multicolumn{1}{c}{$\lambda_{\nu}$} &
\multicolumn{1}{|c|}{$\lambda_{\bar{\nu}}$} &
\multicolumn{1}{c}{$\lambda_{\nu}$} &
\multicolumn{1}{|c|}{$\lambda_{\bar{\nu}}$} &
\multicolumn{1}{c}{$\lambda_{\nu}$} &
\multicolumn{1}{|c|}{$\lambda_{\bar{\nu}}$} \\ \hline

 2.0  &   0.01  &  -100  &  -100  &  -7.99  &  -100  &  -100  &  -100 & -100  &  -100\\
 2.0  &   0.10  &  -64.184  &  -100  &  -8.452  &  -100  &  -100  &  -100  &  -100  &  -45.112\\
 2.0  &   0.20  &  -37.285  &  -100  &  -8.556  &  -93.165  &  -100  &  -100  &  -96.304  &   -29.973\\
 2.0  &   0.40  &  -23.451  &  -100  &  -8.449  &  -48.523  &  -59.425  &  -66.075  &  -52.296  &  -17.72\\
 2.0  &   0.70  &  -17.151  &  -59.608  &  -8.191  &  -28.863  &  -37.187  &  -38.789  &  -33.122  &  -11.164\\
 2.0  &   1.00  &  -13.519  &  -42.43  &  -7.124  &  -21.319  &  -27.246  &  -28.289  &  -24.401  &  -9.072\\
 2.0  &   1.50  &  -10.009  &  -28.721  &  -5.757  &  -15.085  &  -18.942  &  -19.798  &  -17.038  &  -7.477\\
 2.0  &   2.00  &  -8.006  &  -21.691  &  -4.882  &  -11.778  &  -14.604  &  -15.261  &  -13.175  &  -6.467\\
 2.0  &   3.00  &  -5.695  &  -14.497  &  -3.733  &  -8.294  &  -9.986  &  -10.45  &  -9.077  &  -5.122\\
 2.0  &   5.00  &  -3.351  &  -8.409  &  -2.379  &  -5.192  &  -5.745  &  -6.105  &  -5.43  &  -3.493\\
 2.0  &  10.00  &  -0.569  &  -3.009  &  -0.506  &  -1.983  &  -1.522  &  -1.92  &  -1.92  &  -1.182\\
 2.0  &  30.00  &  3.184  &  2.229  &  3.322  &  2.669  &  2.97  &  2.678  &   2.884  &  2.951\\
 5.0  &   0.01  &  -100  &  -100  &  -5.667  &  -100  &  -100  &  -100  &  -100  &  -100\\
 5.0  &   0.10  & -59.658  &  -100  &  -5.669  &  -100  &  -100  &  -100  &  -100  &  -47.691\\
 5.0  &   0.20  &  -33.723  &  -100  &  -5.66  &  -95.628  &  -100  &  -100   &  -92.741  &  -30.924\\
 5.0  &   0.40  &  -20.261  &  -100  &  -5.482  &  -50.659  &  -56.235  &  -69.081  &  -49.106  &  -18.709\\
 5.0  &   0.70  &  -14.13  &  -62.629  &  -5.251  &  -31.31  &  -34.167  &  -41.568  &  -30.101  &  -13.248\\
 5.0  &   1.00  &  -11.459  &  -44.489  &  -5.103  &  -23.227  &  -25.187  &  -30.19  &  -22.341  &  -10.809\\
 5.0  &   1.50  &  -9.119  &  -29.611  &  -4.881  &  -15.966  &  -18.052  &  -20.667  &  -16.148  &  -8.312\\
 5.0  &   2.00  &  -7.683  &  -22.014  &  -4.564  &  -12.099  &  -14.281  &  -15.582  &  -12.851  &  -6.771\\
 5.0  &   3.00  &  -5.636  &  -14.556  &  -3.674  &  -8.352  &  -9.926  &  -10.509  &  -9.017  &  -5.173\\
 5.0  &   5.00  &  -3.342  &  -8.418  &  -2.37  &  -5.201  &  -5.735  &  -6.114  &  -5.421  &  -3.5\\
 5.0  &  10.00  &  -0.568  &  -3.009  &  -0.505  &  -1.984  &  -1.521  &  -1.921  &  -1.918  &  -1.182\\
 5.0  &  30.00  &  3.184  &  2.23  &  3.322  &  2.67  &  2.971  &  2.679  &  2.884  &  2.952\\
 8.0  &   0.01  &  -2.65  &  -100  &  -1.454  &  -100  &  -100  &  -100  &  -100  &  -100\\
 8.0  &   0.10  &  -2.65  &  -100  &  -1.453  &  -100  &  -100  &  -100  &  -84.178  &  -100\\
 8.0  &   0.20  &  -2.645  &  -100  &  -1.444  &  -100  &  -59.392  &  -100  &  -45.153  &  -72.452\\
 8.0  &   0.40  &  -2.626  &  -100  &  -1.348  &  -71.367  &  -32.205  &  -90.891  &  -25.07  &  -39.23\\
 8.0  &   0.70  &  -2.577  &  -74.916  &  -1.201  &  -42.683  &  -20.095  &  -53.541  &  -16.016  &  -24.516\\
 8.0  &   1.00  &  -2.506  &  -53.607  &  -1.112  &  -30.989  &  -15.018  &  -38.342  &  -12.15  &  -18.239\\
 8.0  &   1.50  &  -2.357  &  -36.578  &  -1.005  &  -21.665  &  -10.844  &  -26.353  &  -8.901  &  -13.037\\
 8.0  &   2.00  &  -2.185  &  -27.667  &  -0.908  &  -16.847  &  -8.599  &  -20.229  &  -7.118  &  -10.301\\
 8.0  &   3.00  &  -1.829  &  -18.434  &  -0.732  &  -11.785  &  -6.103  &  -13.86  &  -5.126  &  -7.39\\
 8.0  &   5.00  &  -1.177  &  -10.615  &  -0.45  &  -7.279  &  -3.602  &  -8.167  &  -3.225  &  -4.815\\
 8.0  &  10.00  &  0.161  &  -3.76  &  0.218  &  -2.73  &  -0.786  &  -2.664  &  -1.169  &  -1.891\\
 8.0  &  30.00  &   3.219  &  2.195  &  3.357  &  2.635  &  3.006  &  2.644  &   2.919  &   2.917\\
 11.0  &   0.01  & 5.913  &  -100  &  5.802  &  -100  &  5.731  &  -100  &  5.582  & -100\\
 11.0  &   0.10  & 5.913  &  -100  &  5.798   &  -100  &  5.73  &  -100  &  5.578  &  -100\\
 11.0  &   0.20  &  5.914  &  -100  &  5.801  &  -100  &  5.731  &  -100  &  5.577  &  -100\\
 11.0  &   0.40  &  5.912  &  -100  &  5.802  &  -100  &  5.731  &  -100  &  5.575  &  -100\\
 11.0  &   0.70  &  5.913  &  -100  &  5.807  &  -100  &  5.731  &  -100  &  5.571  &  -100\\
 11.0  &   1.00  &  5.913  &  -100  &  5.811  &  -100  &  5.731  &  -100  &  5.57  &  -100\\
 11.0  &   1.50  &  5.913  &  -100  &  5.816  &  -93.812  &  5.731  &  -98.131  &  5.569  &  -84.642\\
 11.0  &   2.00  &  5.914  &  -81.906  &  5.82  &  -70.997  &  5.732  &  -74.029  &  5.569  &  -64.008\\
 11.0  &   3.00  &  5.915  &  -54.671  &  5.829  &  -47.962  &  5.733  &  -49.728  &  5.573  &  -43.213\\
 11.0  &   5.00  &  5.918  &  -32.505  &  5.844  &  -29.135  &  5.737  &  -29.868  &  5.585   &  -26.383\\
 11.0  &  10.00  &  5.942  &  -15.018  &  5.912  &  -13.984  &  5.77  &  -13.9  &  5.66  &  -13.069\\
 11.0  &  30.00  &  6.322  &  -1.635  &  6.55  &  -1.195  &  6.236  &  -1.185  &  6.285  &  -0.912\\
\end{tabular}}
\end{table}
\begin{table}
\caption{Same as Table~\ref{ta3} but for $^{54}$Cr, $^{55}$Cr,
$^{56}$Cr and $^{57}$Cr} \label{ta4}
{\scriptsize\begin{tabular}{|cc|cccccccc|} $\log\rho \text{Y}_{e}$ &
T$_{9}$ & \multicolumn{2}{c|}{$^{54}$Cr}&
\multicolumn{2}{c|}{$^{55}$Cr} & \multicolumn{2}{c|}{$^{56}$Cr}
& \multicolumn{2}{c|}{$^{57}$Cr}\\
\cline{3-10} & &  \multicolumn{1}{c}{$\lambda_{\nu}$} &
\multicolumn{1}{|c|}{$\lambda_{\bar{\nu}}$} &
\multicolumn{1}{c}{$\lambda_{\nu}$} &
\multicolumn{1}{|c|}{$\lambda_{\bar{\nu}}$} &
\multicolumn{1}{c}{$\lambda_{\nu}$} &
\multicolumn{1}{|c|}{$\lambda_{\bar{\nu}}$} &
\multicolumn{1}{c}{$\lambda_{\nu}$} &
\multicolumn{1}{|c|}{$\lambda_{\bar{\nu}}$} \\ \hline

 2.0  &   0.01  & -100  &  -100 &  -100 &  -2.482  & -100  & -2.762 & -100 &  -1.053\\
 2.0  &   0.10  &  -100  &  -100  &  -100  &  -2.482  & -100  & -2.762 & -100 &  -1.053\\
 2.0  &   0.20  &  -100  &  -63.135  &  -100   &  -2.482  & -100  & -2.762 & -100 &  -1.044\\
 2.0  &   0.40  &  -96.652  & -31.8   &  -82.692  &  -2.482  & -100  & -2.762 & -100 & -0.999 \\
 2.0  &   0.70  & -57.864  &  -18.076  & -49.88   &  -2.464  & -72.849  & -2.762 & -65.044 &  -0.951\\
 2.0  &   1.00  &  -41.311  &  -13.29  & -35.712   &  -2.249  & -51.937  & -2.761 & -46.356 &  -0.927\\
 2.0  &   1.50  &  -27.874  &  -9.768  &  -24.122  &  -1.576  & -35.096  & -2.76 & -31.245 &  -0.907\\
 2.0  &   2.00  &  -20.983  &  -7.9  &  -18.15  &  -1.128  & -26.487  & -2.748 & -23.502 &  -0.892\\
 2.0  &   3.00  &  -13.85  &  -5.797  &  -11.943  &  -0.689  &  -17.593 & -2.635 & -15.502 &  -0.84\\
 2.0  &   5.00  &  -7.731  &  -3.534  &  -6.596  &  -0.3   & -9.921  &  -1.86 & -8.702 &  -0.648\\
 2.0  &  10.00  &  -2.354  &  -0.743  &  -1.91  &  0.695  & -3.162  & 0.04 & -2.906 &  -0.102\\
 2.0  &  30.00  &  2.734  & 3.075   &  2.909  &  3.6  & 2.547  & 3.225 & 2.492 &  2.355\\
 5.0  &   0.01  &  -100  & -100   &  -100 &  -2.501  & -100 & -2.783  & -100 &  -1.054\\
 5.0  &   0.10  &  -100  & -100   &  -100  &  -2.501  & -100  & -2.783 & -100 &  -1.054\\
 5.0  &   0.20  &  -100  &  -66.698  &  -100  &  -2.5  & -100  & -2.781 & -100 &  -1.045\\
 5.0  &   0.40  &  -93.462  & -34.991   &  -79.502  &  -2.497  &   -100  & -2.778 & -100 &  -1\\
 5.0  &   0.70  &  -54.843  &  -21.097  &  -46.859  &  -2.475  & -69.828  & -2.775 & -62.023 &  -0.952\\
 5.0  &   1.00  &  -39.251  &  -15.35  &  -33.652  &  -2.256  & -49.877  & -2.772 & -44.296 &  -0.928\\
 5.0  &   1.50  &  -26.984  &  -10.658  & -23.232   &  -1.58  & -34.206  & -2.77 & -30.355 &  -0.907\\
 5.0  &   2.00  & -20.659   &  -8.224  &  -17.827  &  -1.131  &  -26.163 & -2.761 & -23.178 &  -0.892\\
 5.0  &   3.00  &  -13.791  &  -5.856  & -11.883   &  -0.691  &  -17.533 & -2.652 & -15.442 &  -0.841\\
 5.0  &   5.00  &  -7.722  &  -3.543  &  -6.586  &  -0.302  & -9.912  & -1.867 & -8.692 &  -0.649\\
 5.0  &  10.00  &  -2.353   & -0.744   &  -1.909  &  0.694  & -3.161  &  0.04 &  -2.905 &  -0.102\\
 5.0  &  30.00  &   2.735  &  3.076  &  2.909  &  3.6  & 2.547  & 3.226 & 2.493 & 2.355 \\
 8.0  &   0.01  & -100 &  -100 &  -100 &  -4.442  & -100  & -100 & -100 & -1.381\\
 8.0  &   0.10  &  -100  & -100   &  -100  &  -4.437  & -100  & -27.97 & -100 & -1.379\\
 8.0  &   0.20  &  -100  &  -100  &  -100  &  -4.432   &  -100  & -17.652 &  -100 &  -1.371\\
 8.0  &   0.40  &   -69.429  &  -52.982  & -55.462   & -4.413   & -95.294  & -11.906 & -81.949  &  -1.329 \\
 8.0  &   0.70  & -40.758  & -31.853   &  -32.769  &  -4.27  & -55.76  &  -9.009 & -47.933 &  -1.283  \\
 8.0  &   1.00  &  -29.058   & -23.194   & -23.455   &  -3.656  &  -39.698 & -7.64 & -34.099 & -1.259 \\
 8.0  &   1.50  &  -19.735  &  -16.173  & -15.98   &  -2.712  &  -26.966 & -6.383 & -23.103 &  -1.234\\
 8.0  &   2.00  &  -14.923  &  -12.493  &  -12.088   &  -2.171  & -20.433  & -5.638 & -17.439 & -1.211 \\
 8.0  &   3.00  &  -9.898  &   -8.599 &  -7.986  &  -1.582  &  -13.642 & -4.68 & -11.546 &  -1.129\\
 8.0  &   5.00  &   -5.527  &  -5.078  & -4.386   &  -1.001  & -7.714  & -2.797 & -6.492 & -0.863 \\
 8.0  &  10.00  &  -1.604  &  -1.41  &  -1.157  &  0.072  & -2.409  & -0.4 & -2.153 &  -0.28\\
 8.0  &  30.00  &  2.77  &  3.041  &  2.944  &  3.566  &  2.582  & 3.191 & 2.528 & 2.321 \\
 11.0  &   0.01  &  5.595 &  -100 &  5.494 &  -100  & 4.854  & -100 & 3.966 &  -100\\
 11.0  &   0.10  & 5.59   & -100   & 5.493   &  -100  & 4.858  & -100 & 3.968 &  -100\\
 11.0  &   0.20  &  5.592  &  -100  &  5.49  &  -100  & 4.856  & -100 & 3.969 &  -100\\
 11.0  &   0.40  &  5.593  &  -100  & 5.491   &  -100  &  4.854 & -100 & 3.975 &  -100\\
 11.0  &   0.70  &  5.593 &  -100  &  5.491  &  -100  & 4.854  & -100 & 3.982 &  -100\\
 11.0  &   1.00  & 5.593   &  -100  &  5.492  &  -100  & 4.854  & -100 & 3.986 &  -96.729\\
 11.0  &   1.50  &  5.594  &  -86.624  & 5.503   &  -73.012  & 4.855  &  -76.197 & 3.991 &  -65.243\\
 11.0  &   2.00  &  5.594  &  -65.265  &  5.529  &  -55.054  & 4.855  & -57.373 & 4.005 &  -49.357\\
 11.0  &   3.00  &  5.596  &  -43.68  &  5.593  &  -36.948  & 4.857  & -38.347 & 4.119 &  -33.277\\
 11.0  &   5.00  &  5.601  &  -26.081  & 5.684   &  -22.264   & 4.866  & -22.797 & 4.529 &  -20.139\\
 11.0  &  10.00  &  5.644  & -12.298   &  5.805  &  -10.87  & 5.115  & -10.649 & 5.143 &  -9.963\\
 11.0  &  30.00  &  6.105  &  -0.786  &  6.284  & -0.258   & 5.961  & -0.632 & 5.901 &  -1.489\\

\end{tabular}}
\end{table}

\begin{table}
\caption{Same as Table~\ref{ta3} but for $^{58}$Cr, $^{59}$Cr and
$^{60}$Cr} \label{ta5} {\scriptsize\begin{tabular}{|cc|cccccc|}
$\log\rho \text{Y}_{e}$ & T$_{9}$ & \multicolumn{2}{c|}{$^{58}$Cr}&
\multicolumn{2}{c|}{$^{59}$Cr} & \multicolumn{2}{c|}{$^{60}$Cr} \\
\cline{3-8} & &  \multicolumn{1}{c}{$\lambda_{\nu}$} &
\multicolumn{1}{|c|}{$\lambda_{\bar{\nu}}$} &
\multicolumn{1}{c}{$\lambda_{\nu}$} &
\multicolumn{1}{|c|}{$\lambda_{\bar{\nu}}$} &
\multicolumn{1}{c}{$\lambda_{\nu}$} &
\multicolumn{1}{|c|}{$\lambda_{\bar{\nu}}$} \\ \hline

 2.0  &   0.01  & -100  &  -0.826 &  -100  &  0.396  &  -100 &  0.563\\
 2.0  &   0.10  & -100   &  -0.826  &  -100  &  0.396  & -100  &  0.563\\
 2.0  &   0.20  &  -100  & -0.826   &   -100 &  0.396  &  -100 &  0.563\\
 2.0  &   0.40  &   -100  &   -0.826  &  -100  & 0.395   &   -100 &  0.563\\
 2.0  &   0.70  & -89.901   & -0.826   &  -79.134  &  0.396   & -100  &  0.563\\
 2.0  &   1.00  &  -63.491  &  -0.826  &  -56.211  &  0.423  &  -74.647 &  0.563\\
 2.0  &   1.50  &  -42.335   & -0.826   &  -37.747  &  0.515  & -49.677  &  0.563\\
 2.0  &   2.00  &  -31.552  &  -0.825  &   -28.293  &  0.6  &  -36.989 &  0.563\\
 2.0  &   3.00  &  -20.493  &  -0.815  &  -18.537  &  0.713  &  -24.036 &  0.564\\
 2.0  &   5.00  &  -11.197  &  -0.526  &  -10.273   & 0.837   & -13.259  &  0.662\\
 2.0  &  10.00  &  -3.512  & 1.23   & -3.414   & 1.108   &  -4.519 &  2.061\\
 2.0  &  30.00  & 2.498   & 3.846   & 2.397   &  3.211  & 2.256  &   4.241\\
 5.0  &   0.01  & -100  &  -0.831  &  -100  &  0.395  & -100 &  0.561\\
 5.0  &   0.10  & -100   &  -0.831  & -100   &  0.395  & -100  &  0.561\\
 5.0  &   0.20  &  -100  &   -0.831 &  -100  &  0.395  & -100  &  0.561\\
 5.0  &   0.40  & -100   &  -0.83  &  -100  &  0.395  &  -100 &  0.562\\
 5.0  &   0.70  &  -86.881  & -0.829   & -76.113   &  0.396  & -100  &   0.562\\
 5.0  &   1.00  &  -61.431  &  -0.829  &  -54.151  &  0.423  & -72.587  &  0.562\\
 5.0  &   1.50  &  -41.445  &  -0.828   & -36.857   & 0.514    & -48.787  &  0.562\\
 5.0  &   2.00  &  -31.229  &  -0.828  &  -27.97  &  0.6   & -36.665  &  0.562\\
 5.0  &   3.00  &  -20.433  &   -0.818 &  -18.477  &   0.713  & -23.976  &  0.564\\
 5.0  &   5.00  &  -11.187  &  -0.528   & -10.263   & 0.836   &   -13.25 &  0.661\\
 5.0  &  10.00  &  -3.51  &  1.23  &  -3.413  &  1.108  &  -4.517 &  2.061 \\
 5.0  &  30.00  &  2.499  &  3.846  & 2.398   &  3.212  & 2.257  &   4.241\\
 8.0  &   0.01  &  -100 & -1.82  & -100   &  0.243  & -100  &  0.203\\
 8.0  &   0.10  & -100   & -1.818   & -100   &  0.242  & -100  &  0.202\\
 8.0  &   0.20  &  -100  &  -1.817  &  -100  &  0.241  &  -100 & 0.202 \\
 8.0  &   0.40  &   -100  & -1.813   &  -100  &  0.241  &  -100 &  0.203\\
 8.0  &   0.70  &  -72.79 &  -1.803  & -62.023   &  0.243  &  -88.823 & 0.204 \\
 8.0  &   1.00  &  -51.234  &  -1.788  & -43.954   &  0.274  &  -62.39 &  0.207\\
 8.0  &   1.50  &  -34.193  &  -1.753  & -29.604   &  0.375  & -41.535  & 0.212  \\
 8.0  &   2.00  & -25.489   &  -1.709  &  -22.23  &  0.468  &  -30.926 &  0.22 \\
 8.0  &   3.00  &   -16.536 & -1.604   &   -14.58  & 0.591   & -20.08  &  0.24\\
 8.0  &   5.00  &  -8.987   &  -1.003  & -8.063   &  0.728  &  -11.049 &   0.404\\
 8.0  &  10.00  &  -2.758  &  1.019  &   -2.661  &  0.995  &  -3.765 &   1.952\\
 8.0  &  30.00  &  2.534  & 3.812   &  2.433  &  3.178  &  2.292 &   4.209\\
 11.0  &   0.01  &  3.857  &  -100  & 4.392  &  -100  & 3.364 & -100\\
 11.0  &   0.10  & 3.858   &  -100   & 4.394   & -100   & 3.364  & -100 \\
 11.0  &   0.20  &  3.858  &   -100  &  4.397  &  -100  &   3.364 & -100 \\
 11.0  &   0.40  &  3.857  &  -100  & 4.394   &  -100  &  3.363 &  -100\\
 11.0  &   0.70  &  3.857 &   -100  &  4.389  &  -100  &   3.364 &  -100\\
 11.0  &   1.00  &  3.858  &  -100  &  4.377  &   -83.281  & 3.364  &  -90.844\\
 11.0  &   1.50  & 3.859   &  -67.836  &  4.354  &   -56.099  & 3.366  &  -60.964\\
 11.0  &   2.00  &  3.86   &  -50.968  &  4.334  &   -42.39  & 3.368  &  -45.839\\
 11.0  &   3.00  &  3.864  &  -33.878  & 4.305   &  -28.516   & 3.373  &  -30.447\\
 11.0  &   5.00  &  3.884  &  -19.842  &  4.284  &  -17.163  & 3.399  &  -17.701\\
 11.0  &  10.00  & 4.643   & -8.732   &  4.755  &  -8.254  &  4.155 &  -7.29\\
 11.0  &  30.00  & 5.907  &  0.016  &  5.818  &  -0.609  & 5.755  &  0.539\\

\end{tabular}}
\end{table}

\end{document}